\def\wh{wormhole }
\def\beq{\begin{equation}}
\def\eeq{\end{equation}}
\def\bea{\begin{eqnarray}}
\def\eea{\end{eqnarray}}
\def\ep{\epsilon}
\def\om{\omega}
\def\th{_{_{\rm th}}}
\documentstyle[prl,preprint,aps]{revtex}

\begin{document}
\title{Brans-Dicke wormholes in nonvacuum spacetime}
\author{Luis A. Anchordoqui\thanks{E-mail: doqui@venus.fisica.unlp.edu.ar}, 
Santiago Perez Bergliaffa \& Diego F. Torres}

\address{Departamento de F\'{\i}sica, Universidad Nacional de La Plata\\
C.C. 67, 1900, La Plata, Buenos Aires, Argentina}
\maketitle
%\thispagestyle{empty}
%\def\thepage{\protect\raisebox{0ex}{\ } DRAFT - 1.3}
%\thispagestyle{headings}
%\markright{\thepage}

\begin{abstract}
Analytical wormhole solutions in Brans-Dicke theory in the presence of matter
are presented. It is shown that 
the \wh throat must not be necessarily
threaded with exotic matter.

\noindent PACS {\em number(s):} 04.20Jb, 04.50+h
\end{abstract}

\newpage
%\pagenumbering{arabic}
The field equations of general relativity, being local in character, admit 
solutions with nontrivial topology. Among these, wormholes have been 
extensively studied \cite{motho}. Their most salient feature is that an 
embedding of one of their spacelike sections in Euclidean space displays two 
asymptotically flat regions joined by a throat.

The interest on wormholes is twofold. From the point of view of the Euclidean 
path integral formulation of quantum gravity, Coleman \cite{col1} and Giddings
and Strominger \cite{gistro}, among others, have shown that the effect of
wormholes is to modify low energy coupling constants and to provide 
probability distributions for them. In particular, Coleman \cite{col2} showed 
that, in the dilute wormhole approximation, the probability distribution for 
universes is infinitely peaked at $\Lambda = 0$, rendering all other values of
the cosmological constant improbable.

On the purely gravitational side, the interest 
has been recently focused on traversable \wh 
\cite{motho,mothoyu,viss,nontra,fuwhee}. Most of the efforts are directed to study static 
configurations \cite{kar} that must have a number of 
specific properties in order to be traversable. The
most striking of these properties is the violation of the energy 
conditions \cite{wald}. 
It implies that the matter that generates the 
\wh is exotic \cite{motho}, {\em viz.} its energy density is negative, 
as seen by static 
observers. Geometrically, this is a direct consequence of the 
singularity
theorems of Hawking and Penrose \cite{hawpen}. Although we do not know of any 
such exotic
material to date, quantum field theory might come to the rescue \cite{edneg}.

Finally, we should mention yet another proposal related to wormholes. It has 
been shown \cite{mothoyu,frimo} that a nonstatic wormhole's throat can be 
transformed into a time tunnel. Physical effects in this type of spacetimes 
have been studied in \cite{fronov}.

Wormhole solutions have also been discussed in alternative 
theories of gravity, such as $R + R^2$ theories \cite{hoch}, Moffat's 
nonsymmetric theory
\cite{mof}, Einstein-Gauss-Bonnet theory \cite{bha}, and Brans-Dicke (BD) 
theory
\cite{agn}. In the last case,
static \wh solutions were found in vacuum, the source of gravity being the 
scalar field. Dynamical 
solutions are discussed in \cite{acce}. The aim of this paper is to 
look for static \wh solutions of
Brans-Dicke theory in a general setting, {\em i.e.} in the presence of matter 
that obeys a generic equation of state \cite{Let}. We shall also discuss
whether the BD scalar can be the ``carrier'' of exoticity, as was shown
in \cite{agn} for the vacuum case.

Following the conventions of \cite{bruck}, the field equations of Brans-Dicke 
theory are
\jot= 17pt
\begin{eqnarray}
R_{\mu\nu} & = & {8\pi\over\Phi}\left( T_{\mu\nu}- {\omega +1\over 2
\omega + 3}\: T\: g_{\mu\nu}\right) + \omega\: {\Phi_{;\mu}\Phi_{;\nu}
\over\Phi^2}
+ {\Phi_{;\mu;\nu}\over\Phi} \\
\Phi^{;\mu}_{\; ;\mu} & = & {8\pi\over 2\omega +3}\: T
\end{eqnarray}

The assumption of a static spacetime entails that it is possible to choose
a metric and a scalar field such that
\beq 
g_{\mu\nu ,t} = 0 \;\;\;\;\;\;\;\;\;\;  \Phi _{,t} = 0 
\;\;\;\;\;\;\;\;\;\; g_{ti} = 0 
\eeq
($i=r,\theta,\phi$). We further require spherical symmetry, so that the line 
element can be written in Schwarzschild form:
\beq
ds^2 = -e^{2\psi}dt^2 + e^{2\lambda} dr^2 + r^2 ( d\theta^2 + \sin^2\theta
d\phi ^2)
\label{met}
\eeq
For the stress-energy tensor of matter we choose
\beq
T^t_{\;t} =- \rho (r)\;\;\;\;\;\;\;\;\;\;  T^r_{\;r} = -\tau (r) \;\;
\;\;\;\;\;\;\;\;  T^\theta_{\;\theta} = T^\phi_{\;\phi} = p(r)
\eeq
and zero otherwise. Finally, we adopt the following equation of state for 
matter:
\beq
-\tau + 2p = \epsilon \rho
\label{stateq}
\eeq
where $\epsilon$ is a constant. Now, the trace of the stress-energy tensor 
can be written as $T = -\tau + 2p - \rho = \rho (\epsilon -1)$. 
The field equations take the form
\jot=17pt
\begin{mathletters}
\beq
-\psi''- (\psi')^2 + \lambda'\psi'+ 2 {\lambda'\over r} 
=  -{8\pi\over\Phi}\left[ \tau +{\omega +1\over 2\omega +3}\:T\right]
e^{2\lambda}+(\omega +1) (\ln\Phi)'^2 
+ (\ln\Phi)''-\lambda' (\ln\Phi)' 
\label{fieq1}
\eeq
\beq
1 - 
r e^{-2\lambda}\left[\psi ' - \lambda ' + {1\over r}\right]  =  
{8\pi\over\Phi}\left[ p - {\omega + 1\over 2\omega +3}\:T\right]r^2 + r 
e^{-2\lambda}(\ln \Phi)' 
\label{fieq2}
\eeq
\beq
e^{2(\psi -\lambda)}\left[\psi '' + (\psi ')^2 - \lambda ' \psi ' + 
2 {\psi '
\over r}\right]  =  {8\pi\over\Phi}\left[\rho + {\omega + 1\over 2\omega +3}
\: T\right] e^{2\psi} - \psi ' e^{2(\psi - \lambda)} (\ln \Phi)' 
\label{fieq3}
\eeq
\beq
\Phi '' - \Phi ' \left( \lambda ' - \psi ' - {2\over r}\right)  =  
{8\pi\over 2\omega + 3}\: T \: e^{2\lambda}
\label{fieq4}
\eeq
\end{mathletters}
To solve the system made up of Eqs. (7) we shall follow the 
philosophy sketched in 
\cite{karsah}. We shall look for a differential equation relating $\psi$ and 
$\lambda$, starting from the equations of motion and the equation of state. 
The 
equation we shall obtain is second order and nonlinear in $\psi$ 
but, after a change of variables, first order and linear in $\lambda$. 
We shall
then make a specific choice for $\psi$ consistent with asymptotic flatness
and nonexistence of horizons and singularities. We shall finally 
substitute this 
$\psi$ into the linear equation and solve for $\lambda$. 

As explained in \cite{bruck}, from Eqs. (\ref{stateq}), (\ref{fieq3}), and 
(\ref{fieq4}), it can be shown that
$
\Phi = \Phi _0\, e^{c\,\psi}
$
where $c = (\epsilon - 1)/[2\omega +3 + (\omega +1)(\epsilon -1)]$, and
$\Phi_0$ is related to the value of the gravitational coupling
constant when $r \rightarrow \infty$. 
In the case $\omega\rightarrow\infty$ or $\epsilon
\rightarrow 1$, we get general relativity back (although in the latter case,
other solutions different from $\Phi = $ const might exist).

After a bit of algebra, we get the equation:
\beq
A\,\psi '' + B\, (\psi ')^2 + 2A\, \psi ' - A \,\lambda ' \psi ' + 
{2\over r^2}\: 
(e^{2\lambda} - 1 ) = 0
\label{cons}
\eeq
where
\jot=17pt
\begin{displaymath}
A =  -2 \,\, {2+\ep + 2 \om\over 2+\ep +\om (1+\ep)}
\end{displaymath}
\begin{displaymath} 
B  =  -\; {8 + \ep^2(\om +2) + 4\om^2 (1+\ep) +8\ep + 11\om + 12 \om\ep\over
[2+\ep + \om (\ep +1)]^2} 
\end{displaymath}
In the spirit of \cite{karsah}, we make the ansatz $\psi = - \alpha / r$, 
where $\alpha$ is a positive constant. 
With this election, which guarantees that the gravitational constant takes the
correct value at $r\rightarrow\infty$,
Eq. (\ref{cons}) takes the form
\beq
h(r) + f(r)\; e^{2\lambda} + g(r)\;\lambda ' =0
\label{difeq}
\eeq
where
\begin{displaymath}
h(r)  = B\left({\alpha\over r^2}\right)^2- {2\over r^2} 
\;\;\;\;\;\;\;\;\;\;\;\;
f(r)  =  {2\over r^2} \;\;\;\;\;\;\;\;\;\;\;\; 
g(r) =  -{A\alpha\over r^2} + {4\over r}
\end{displaymath}
A suitable change of variables transforms Eq. (\ref{difeq}) 
into a Bernoulli equation, and afterwards into a linear equation. Its general 
solution is given by
\beq
e^{-2\lambda} = {e^{2s/\varphi}\over \varphi}\left(1+{R\over\varphi}\right)
^{-(8l+1)}\left\{I +{\cal K}\right\}
\label{gensol}
\eeq
where
\jot=17pt
\begin{displaymath}
\varphi = {r\over\alpha} \;\;\;\;\;\;\;\;\;\; s = {B\over A}  
\;\;\;\;\;\;\;\;\;\; R = -{A\over 4} \;\;\;\;\;\;\;\;\;\; 
l = -{B\over A^2} 
\end{displaymath}
\begin{displaymath}
I \equiv \int e^{-2s/\varphi}\left( 1+{R\over\varphi}\right)^{8l}d\varphi
\end{displaymath}
and ${\cal K}$ is a constant. It is not valid when $A \rightarrow 0$, 
{\em i.e.} for $\omega = -1-\ep/2$. 
The binomial $(1 +R/\varphi)^{8l}$ is related 
to the hypergeometric function
$_2F_1$ \cite{erdel}. Using the relation \cite{erdel}
\beq
e^t\:{_pF_q}(\alpha_1,\ldots\alpha_p;\beta_1\ldots\beta_q;-xt) = \sum_{n=0}
^{\infty}\;{_{p+1}F_q}(-n,\alpha_1,\ldots\alpha_p;\beta_1,\ldots\beta_q;x)
{t^n\over n!}\:,
\eeq
the integral $I$ can be written
\beq
I = 2s\sum_{n=0}^{\infty}\int {_3F_1}(-n,-8l,b;b;R/2s)\:\left({-2s\over\varphi}
\right)^n
\eeq
Integrating out the terms corresponding to $n=0$ and $n=1$, we finally get
\beq
I = \varphi - 8\,l\, R\, \ln\varphi + \varphi
\sum_{n=2}^\infty \; {_3F_1}(-n,8l,b;b;R/2s)(-1)^{n}\left({2s\over\varphi}
\right)^n
{1\over n! \, (n-1)} 
\eeq
It is easily seen that $e^{2\lambda} \rightarrow  1$ when $\varphi
\rightarrow \infty$.

In order to fix the constant ${\cal K}$, we must select a value for the 
dimensionless radius ($\varphi\th$) such that 
 the ``flaring out'' 
condition
\beq
\lim_{\varphi \rightarrow \varphi\th^{^+}} e^{-2\lambda} = 0^+ 
\label{fo}
\eeq
is satisfied. In the case $R \leq 0$, $\varphi\th$ must necessarily 
be greater than $|R|$, so that the flaring
out condition
holds for all values of $\om$ and $\ep$ except, obviously, those where $R$ 
diverges, which are given by $\om = -(2+\ep)/(1+\ep)$. 
Nevertheless, the absolute size of the throat also depends on $\alpha$
\footnote{This situation is analogous to what Kar and Sahdev have found
for wormholes in general relativity \cite{karsah}.}. 
The aforementioned properties of $\lambda$, together with the definition 
of $\psi$, bear out that 
the metric tensor describes two asymptotically flat spacetimes joined by a 
throat. 

Let us now study the issue of weak energy condition (WEC) violation.
Using the field equations and the expression for the trace, we easily obtain 
\beq
{2e^{2\lambda}\over r^2} - {4\psi '\over r} - {2\over r^2} = {16\pi\over\Phi}
\tau e^{2\lambda} + {4\over r}{\Phi '\over\Phi} - \om\left({\Phi '
\over\Phi}
\right)^2 + 2 {\Phi '\over\Phi}\psi '
\label{tau}
\eeq
At the throat, $e^{2\lambda}\rightarrow\infty$, and then
\beq
\tau\th \approx{\Phi\th \over 8\pi r\th^2}
\label{tauth}
\eeq
To calculate $\rho\th$, we use the nontrivial component of the 
equation $ T^{\mu}_{\;\nu;\mu}=0$:
\beq
\tau'  = \psi ' (\rho - \tau ) - {2\tau\over r} - 
{\epsilon\rho + \tau\over r}
\label{econ}
\eeq
Using Eqs. (\ref{tauth}) and (\ref{econ}), and the derivative of Eq.
(\ref{tau}),
\beq
\rho\th \approx \tau\th\; 
{c+1+ \varphi\th \over 1-\ep\varphi\th}
\label{roth}
\eeq
And finally, from Eq. (\ref{stateq}),   
\beq
p\th \approx {\tau\th\over 2} \; {\ep\,
(c+1) + 1 \over   1 - \ep\,\varphi\th}
\label{pth}
\eeq
We shall show now that WEC may be violated (at least near the throat) with 
nonexotic matter. This means that we shall present the parameters for which
a \wh solution exists whenever
the matter content of the theory satisfying
the inequalities
\beq
\rho\th\geq 0\;\;\;\;\;\;\;\;\;
\rho\th - \tau\th \geq 0\;\;\;\;\;\;\;\;\;
\rho\th + p\th \geq 0
\label{matwec}
\eeq  
or equivalently,
\bea
{c+1+ \varphi\th \over 1-\ep\varphi\th} & \geq & 1 \label{read1}\\
{\ep (c+1) + 3 + 2 (c +\varphi\th) \over 1-\ep\varphi\th} & \geq  & 0 
\label{read2}
\eea
In addition, a necessary condition for the violation of the weak energy 
condition for matter plus Brans-Dicke field at the throat is given by 
\bea
{2(\om +1)+\ep\over 2\om +3} \; \rho\th & \leq & 0 \label{d1}
\eea

As an example, let us study the case $\ep = 2$. 
From Eqs. (\ref{tauth}), (\ref{roth}), and (\ref{pth}), the inequalities
(\ref{matwec}) will be satisfied if
\beq
\left(\varphi\th\geq -{1\over 9\,\om +12}\;\;\;{\rm  and}\;\;\;
 \varphi\th < {1\over 2} 
\right)\;\;\;\;{\rm or}\;\;\;\;
\left(\varphi\th\leq -{1\over 9\,\om +12}\;\;\; {\rm and }\;\;\;
 \varphi\th > {1\over 2}\right)
\label{twocon}
\eeq
Inequality (\ref{d1}) will be satisfied for $\om \in (-2,-3/2)$. Finally, 
we have to impose that
$\varphi\th\geq |A/4|$, which implies that 
\beq
\varphi\th\geq \left| {2+\om\over4+3\om} \right|
\label{ult}
\eeq
These inequalities constrain $\varphi\th$
to an interval in which 
a nonexotic wormhole can be 
constructed, for instance, in the case $\om = -1.75$. We should 
recall that a definite interval for $\varphi\th$ does not determine 
the radius of the throat, because of the dependence of $\varphi$ on 
$\alpha$.

Summing up, we showed that Brans-Dicke theory in the presence of 
matter with a fairly general equation of state admits analytical wormhole 
solutions.
They generalize the vacuum ones presented by Agnese and La Camera \cite{agn}.
It should be noted that there exists some regions of the parameter space
in which the Brans-Dicke field may play the role of exotic 
matter, implying that 
it might be possible to build a {\em wormholelike} spacetime with the 
presence of ordinary matter at the throat.

\vspace{1.5cm}

We are grateful to H. Vucetich for valuable
discussions and A. Newman for inestimable inspiration. 
L.A.A. held partial support from FOMEC. S.P.B. and D.F.T.
were partially supported by CONICET. S.P.B. also acknowledges  
partial support by UNLP.


\begin{thebibliography}{99}
\bibitem{motho} M. Morris and K. Thorne, Am. J. Phys. {\bf 56}, 395 (1988), and
references therein.
\bibitem{col1} S. Coleman, Nuc. Phys. {\bf B307}, 867 (1988)
\bibitem{gistro} S. Giddings and A. Strominger, Nuc. Phys. {\bf B321}, 481 (1988).
\bibitem{col2} S. Coleman, Nuc. Phys. {\bf B310}, 643 (1988).
\bibitem{mothoyu} M. Morris, K. Thorne and U. Yurtserver, Phys. Rev. Lett.
{\bf 61}, 1446 (1988).
\bibitem{viss} M. Visser, Phys. Rev. D {\bf 39}, 3182 (1989).
\bibitem{nontra} A classical example of  nontraversable \wh is the 
Schwarzschild \wh, which ``pinches off'' before any signal can travel through
it, yielding two singularities \cite{fuwhee}.
\bibitem{fuwhee} R. Fuller and J. Wheeler, Phys. Rev. D {\bf 128}, 9191 (1962).
\bibitem{kar} Nonstatic wormholes that do not require WEC violating matter
at least for a finite interval of time
have been studied in S. Kar, Phys. Rev. D {\bf 49}, 862 (1994) and 
A. Wang and P. Letelier, Prog. Theor. Phys. {\bf 94} L137 (1995)
\bibitem{wald} R. M. Wald, {\it General Relativity} (University 
of Chicago Press,
Chicago, 1984).
\bibitem{hawpen} S. Hawking and G. Ellis, {\it The large scale structure of
space time}, (Cambridge University Press, Cambridge, England, 1973)
\bibitem{edneg} H. Epstein, V. Glasser, and A. Jaffe, Nuovo Cimento {\bf 36}, 
2296 (1965);
L. Parker and S. Fulling, Phys. Rev. D {\bf 7}, 2357 (1973);
L. Ford, Proc. R. Soc. London, {\bf A364}, 227 (1978).
\bibitem{frimo} J. Friedman, M. Morris, I. Novikov, F. Echeverr\'{\i}a, G. 
Klinkhammer, K. Thorne, and U. Yurtserver, Phys. Rev. D {\bf 42}, 1915 (1990).
\bibitem{fronov} V. Frolov and I. Novikov, Phys. Rev. D {\bf 42}, 1057 (1990).
\bibitem{hoch} D. Hochberg, Phys. Lett. B {\bf 251}, 349 (1990).
\bibitem{mof} J. Moffat and T. Svodoba, Phys. Rev. D {\bf 44}, 429 (1991).
\bibitem{bha} B. Bhawal and S. Kar, Phys. Rev. D {\bf 46}, 2464 (1992).
\bibitem{agn} A. Agnese and M. La Camera, Phys. Rev. D {\bf 51}, 2011 (1995).
\bibitem{acce} F. Accetta, A. Chodos and B. Shao, Nuc. Phys. {\bf B333}, 221
(1990).
\bibitem{Let} P. S. Letelier and A. Wang 
have studied some wormhole solutions in Brans-Dicke theory from the point of view 
of bubbles in Phys. Rev. D {\bf 48}, 631 (1993).
\bibitem{bruck} W. Bruckman and E. Kazes, Phys. Rev. D {\bf 16}, 261 (1977).
\bibitem{karsah} S. Kar and D. Sahdev, Phys. Rev D {\bf 52}, 2030 (1995).
\bibitem{erdel} {\it Higher transcendental functions}, 
(Bateman Manuscript Project), edited by A. Erd\'elyi {\it et al.} 
(McGraw-Hill, New York, 1955),
Vol. I, p. 101, and Vol. III, p.267.
\end{thebibliography}
\end{document}